\documentclass[twoside]{article}
\usepackage{landau}

\usepackage{amsmath}
  \usepackage{paralist}
  \usepackage{epsfig} 

\usepackage[latin1]{inputenc}
\usepackage{amssymb,amsmath}
\usepackage{latexsym}
\usepackage{fancyhdr}
\usepackage{color}
\usepackage{varioref,xr-hyper}

\usepackage[colorlinks=true, citecolor=red]{hyperref}
 \usepackage{amsfonts}       
\usepackage{tikz}
\usepackage{graphicx}        
\usepackage[bottom]{footmisc}
\usepackage{latexsym,epic,eepic,epsfig}
\usepackage{pdfsync}

\DeclareGraphicsExtensions{.pdf, .eps, .jpg, .tif}
\def\eq#1{ {\begin{eqnarray}#1\end{eqnarray}}}
\def\n{\nabla}

\def\p{\partial}

\def\ds{\displaystyle}
\def\tr{\hbox{tr}}
\def\det{\hbox{det}}

\def\D{{\mathrm{D}}}
\def\Dt{{\mathbb{D}_t}}
\def\Y{{\mathbb{X}}}
\def\Y{{\mathbb{Y}}}

\def\D{{\mathrm{D}}}

\def\R{{\mathbb R}}

\def\T{{\mathcal T}}

\def\vd{{\mathbf d}}
\def\vu{{\mathbf u}}
\def\vv{{\mathbf v}}
\def\vf{{\mathbf f}}
\def\vn{{\mathbf n}}
\def\vE{{\mathbf E}}
\def\vF{{\mathbf F}}
\def\vH{{\mathbf H}}
\def\vB{{\mathbf B}}
\def\vI{{\mathbf I}}
\def\vA{{\mathbf A}}
\def\vB{{\mathbf B}}
\def\vC{{\mathbf C}}
\def\vX{{\mathbf X}}
\def\vone{{\mathbf 1}}

\newtheorem{theorem}{Theorem}[section]

\newtheorem{lemma}[theorem]{Lemma}
\newtheorem{proposition}{Proposition}

\theoremstyle{definition}

\newtheorem{remark}{Remark}

\def\shorttitle{An Energy Stable Monolithic FSI scheme}
\def\shortauthor{O. Pironneau}

\begin{document}

\title
{\Large \bf \boldmath An Energy Stable Monolithic Eulerian Fluid-Structure Numerical Scheme
\footnote{Written in honour of Philippe Ciarlet for his $80^{th}$ birthday.}}%

\author{
    \large Olivier Pironneau
    \\ \normalsize\emph{Sorbonne Universit\'e, UPMC (Paris VI)}
    \\ \normalsize\emph{ Laboratoire Jacques-Louis Lions}
    \\ \normalsize\emph{Place Jussieu, Boite 187, Paris 75252, France}
    \\ \normalsize\emph{E-mail: Olivier.Pironneau@upmc.fr}
}

\date{{\footnotesize Submitted in July 2016 to ``Chinese Annals of Mathematics", 
\\
Herv\'e Ledret, Annie Raoult, Tatsien Li (eds). 2017. }}

\maketitle

\thispagestyle{first}


\abstract{
The conservation laws of continuum mechanics,  written in an Eulerian frame, do not distinguish fluids and solids, except in the expression of the stress tensors, usually with Newton's hypothesis for the fluids and Helmholtz potentials of energy for hyperelastic solids. By taking the velocities as unknown monolithic methods for fluid structure interactions (FSI) are built.  In this article such a formulation is analysed when the fluid is compressible and the fluid is incompressible.
 The idea is not new but the progress of mesh generators and numerical schemes like the Characteristics-Galerkin method render this approach feasible and reasonably robust.
In this article the method and its discretisation are presented, stability is discussed through an energy estimate.  
A numerical section discusses implementation issues and presents a few simple tests.
}


\section*{Introduction}
Currently two methods dominate FSI (Fluid-Structure-Interaction) science: Arbitrary Lagrangian Eulerian (ALE) methods especially for thin structures \cite{nobile1}\cite{quarteroni} and immersed boundary methods (IBM)\cite{peskin}\cite{coupez}, for which the mathematical analysis is more advanced\cite{gastaldi} but the numerical implementations lag behind. ALE for large displacements have meshing difficulties \cite{jliu} and to a lesser extent with the matching conditions at the fluid-solid interface\cite{letallechauret}. Furthermore, iterative solvers for ALE-based FSI methods which rely on alternative solutions of the fluid and the structure parts are subject to the added mass effect and require special solvers\cite{fernandez}\cite{canic}.  

Alternatives to ALE and  IBM are few. One old method \cite{bathe}\cite{bathe2} has resurfaced recently,  the so-called \emph{actualized Lagrangian methods} for computing structures \cite{leger} \cite{liu} (see also \cite{cottet} although different from the present study because it deals mostly with membranes).
\\\\
Continuum mechanics doesn't distinguish between solids and fluids till it comes to the constitutive equations.  This has been exploited numerically in several studies but most often in the context of ALE\cite{letallec}\cite{turek}\cite{richter}. 

In the present study, which is a follow-up of \cite{erice} and \cite{fhopfsi}, we investigate what Stephan Turek\cite{turek} Heil\cite{heil} and Wang\cite{wang} call a monolithic formulation but here in an Eulerian framework, as in \cite{dunne}\cite{dunne2}\cite{rannacher}\cite{dunnerannacher}, following the displaced geometry of the fluid and the solid.  
In \cite{dunne}, the authors obtained excellent results with the fully Eulerian formulation adopted here but at the cost of meshing difficulties to handle the Lagrangian derivatives. Here we advocated the Characteristic-Galerkin method and obtain an energy estimate, which is not a proof of stability but a prerequisite for it.

\section{Conservation Laws}
Let the time dependent computational domain $\Omega_t$ be made of a fluid  region $\Omega_t^f$ and a solid  region $\Omega_t^s$ with no overlap: $\overline\Omega_t=\overline\Omega_t^f\cup\overline\Omega_t^s$,  $\Omega_t^f\cap\Omega_t^s=\emptyset$ at any times $t\in(0,T)$. At initial time $\Omega_0^f$ and $\Omega_0^s$ are prescribed. 

Let the fluid-structure interface be $\Sigma_t=\overline\Omega_t^f\cap\overline\Omega_t^s$ and the boundary  of $\Omega_t$ be $\p\Omega_t$. The part of $\p\Omega_t$ on which either the structure is clamped or on which there is a no slip condition on the fluid, that part is denoted by $\Gamma$ and assumed to be independent of time.

 The following standard notations are used. For more details  see one of a textbook: \cite{ciarlet},\cite{hughes},\cite{bathe},\cite{antman}, or the following article: ,\cite{turek},\cite{letallec}.
For clarity we use bold characters for vectors and tensors/matrices, with some exceptions, like $x,x^0\in\R^d, d=2$ or 3.
\begin{itemize}
\item $\vX:\Omega_0\times(0,T)\mapsto\Omega_t$: $\vX(x^0,t)$, the Lagrangian position at t of $x^0$.
\item $\vu=\p_t\vX$, the velocity of the deformation,
\item $\vF=\n^T\vX=((\p_{x^0_i}\vX_j))$, the Jacobian of the deformation,
\item $J=\det_\vF$. 
\end{itemize}
We denote by $\tr_A$ and $\det_A$ the trace and determinant of $A$.  
To describe the fluid structure system we need the following:
\begin{itemize}
\item  $\rho=\vone_{\Omega^f_t}\rho^f+\vone_{\Omega^s_t}\rho^s$, the density,
\item $\sigma=\vone_{\Omega^f_t}\sigma^f+\vone_{\Omega^s_t}\sigma^s$, the stress tensor,
\item $\vf(x,t)$ the density of volumic forces at $x,t$.
\item $\vd=\vX(x^0,t)-x^0$, the displacement.
\end{itemize}
Finally and unless specified all spatial derivatives are with respect to $x\in\Omega_t$ and not with respect to $x^0\in\Omega_0$.  Let $\phi$ a function of $x,t$; as $x=\vX(x^0,t),~x^0\in\Omega_0$, $\phi$ is also a function of $x^0$ and we have: 
\[
\n_{x^0} \phi =[\p_{x^0_i} \phi] =[\p_{x^0_i}{\vX_j}\p_{x_j}\phi] =\vF^T\n\phi.
\]
When $\vX$ is one-to-one and invertible, $\vd$ and $\vF$ can be seen as functions of $(x,t)$ instead of $(x^0,t)$.  They are related by
\[
\vF^T=\n_{x^0}\vX=\n_{x^0}(\vd+x^0) = \n_{x^0}\vd+\vI=\vF^T\n\vd+\vI,~~~ \Rightarrow~~\vF=(\vI-\n\vd)^{-T}
\]
Time derivatives are related by (note the notation $\Dt$)
\[
 \Dt\phi:=\frac d{dt} \phi(\vX(x_0,t),t)_{|x=\vX(x_0,t)} =\p_t \phi(x,t)+ \vu\cdot\n \phi(x,t).
\]
It is convenient to introduce  (note the difference between $\Dt$ above and $\D$ here):
\[
\D\vu=\n\vu+\n^T\vu.
\]
Conservation of momentum and conservation of mass take the same form for the fluid and the solid:
\[ \ds \rho \Dt\vu=\vf+ \n\cdot\sigma,~~~ ~ \Dt\rho+\rho\n\cdot\vu=\Dt(J\rho)=0,
\]
So $J\rho=\rho_0$ at all times and
\eq{
J^{-1}\rho_0 \Dt\vu=f+ \n\cdot\sigma\hbox{ in }\Omega_t,~\forall t\in(0,T),
}
with continuity of $\vu$ and of $\sigma\cdot\vn$ at the fluid-structure interface $\Sigma$ in absence of interface constraint like surface tension.
There are also  unwritten constraints pertaining to the realisability of the map $\vX$ (see \cite{ciarlet},\cite{hughes}).
%

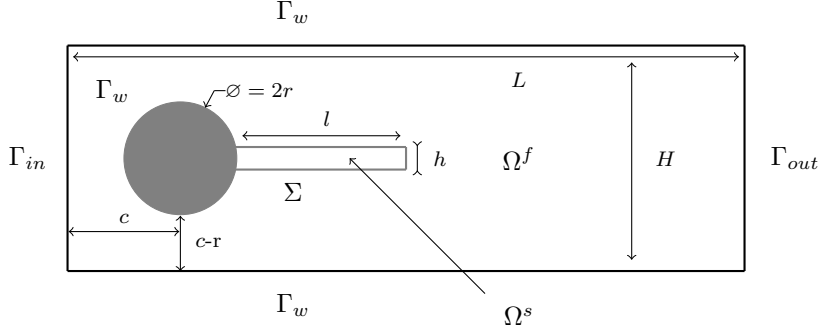
\begin{figure}
\begin{center}
\begin{tikzpicture}[xscale=1.5,yscale=1.5]
\draw [gray,fill] (0,0) circle [radius=0.5];
\draw [thick] (-1,-1) -- (5,-1);
\draw [thick] (-1,1) -- (5,1);
\draw [thick] (-1,-1) -- (-1,1);
\draw [thick] (5,-1) -- (5,1);
\draw [gray,thick] (0.4,0.1) -- (2,0.1);
\draw [gray,thick] (0.4,-0.1) -- (2,-0.1);
\draw [gray,thick] (2,-0.1) -- (2,0.1);
\draw [<->] (-0.95,0.9) -- (4.95,0.9);
\node at (3,0.7) {{\footnotesize$L$}};
\draw [<->] (4,-0.9) -- (4,0.85);
\node at (4.3,0.) {{\footnotesize$H$}};
\draw [>-<] (2.1,-0.15) -- (2.1,0.15);
\node at (2.3,0.) {{\footnotesize$h$}};
\draw [<->] (0.55,0.2) -- (1.95,0.2);
\node at (1.3,0.35) {{\footnotesize$l$}};
\draw [<->] (0.,-0.51) -- (0.,-1);
\node at (0.25,-0.75) {{\footnotesize$c$-r}};
\draw [<->] (-0.99,-0.65) -- (-0.01,-0.65);
\node at (-0.5,-0.53) {{\footnotesize$c$}};
\draw [->] (0.3,0.6) -- (0.22,0.45);
\draw [-] (0.3,0.6) -- (0.4,0.6);
\node at (0.7,0.6) {{\footnotesize$\varnothing=2r$}};
\node at (-1.35,0) {$\Gamma_{in}$};
\node at (5.45,0) {$\Gamma_{out}$};
\node at (1,-1.3) {$\Gamma_w$};
\node at (1,1.3) {$\Gamma_w$};
\node at (-0.6,0.6) {$\Gamma_w$};
\node at (1,-0.3) {$\Sigma$};
\node at (3,0.) {$\Omega^f$};
\node at (3,-1.4) {$\Omega^s$};
\draw [->] (2.7,-1.2)--(1.5,0);
\end{tikzpicture}
\caption{\label{sketch}{\it The geometry of the FLUSTRUK test\cite{dunnerannacher}. The cylinder (in black) is fixed but the flag is a thick compressible Mooney-Rivlin material clamped to the cylinder by its left boundary; the outer rectangle is filled with a fluid which enters from the left $\Gamma_{in}$ and leaves on the right $\Gamma_{out}$; the horizontal boundaries of the outer rectangle are walls, so they form together with the cylinder the boundary $\Gamma_w$.  The flag is at time zero a rectangle of size $l\times h$.  The outer rectangle has size $L\times H$.  The center of the circle representing the cylinder is at $(c,c)$ in a frame of reference which has the lower left corner at $(0,0)$; the cylinder has radius $r$ and is fixed.}}
\end{center}
\end{figure}

\subsection{Constitutive Equations}  
We consider a bi-diemsional geometry. For the 3d case, see \cite{CYOP}.
\begin{itemize}
\item For a  Newtonian incompressible fluid :
$
\sigma^f=-p^f{\bf I} +\mu^f\D\vu
$
\item  For an hyperelastic material :
$
\sigma^s= \rho^s \p_{\vF}\Psi\vF^T
$ 
\end{itemize}
where $\Psi$ is the Helmholtz potential which, in the case of a S$^t$-Venant-Kirchhoff material, is \cite{ciarlet} 
\eq{\label{defpsi}
\Psi(\vF)= \frac{\lambda^s}2\tr^2_{\vE}+\mu^s\tr_{\vE^2},~~\vE=\frac12(\vF^T\vF-\vI)
}
It is easy to see that $\tr_\vE=\frac12\tr_{\vF^T\vF}-1$ and
\eq{&
\p_\vF\tr_{ \vF^T\vF}&=((\p_{\vF_{ij}}\sum_{m,n}F_{m,n}^2))=2\vF
~~\Rightarrow~\p_\vF\tr_\vE=\vF
\cr&
\p_\vF\tr_{(\vF^T\vF)^2}&=((\p_{\vF_{ij}}\sum_{n,m,p,k}F_{n,k}F_{n,m}F_{p,m}F_{p,k}))=4\vF\vF^T\vF
}
which implies that $\p_\vF\tr_{\vE^2}= 2\vF\vE$.  Therefore
  \[
  \p_\vF\Psi(\vF)\vF^T=(\lambda^s\tr_\vE\vF+2\mu^s\vF\vE)\vF^T
\]
which in turn implies that
\[
 \sigma^s= \rho^s \vF(\lambda^s\tr_\vE+2\mu^s\vE)\vF^T
 = J^{-1}\rho^s_0 \vF(\lambda^s\tr_\vE+2\mu^s\vE)\vF^T
\]

For a tensor $\vA$ define $|\vA|=\sum_{ij}A^2_{ij}$.
\begin{remark}
Some authors have a different definition for the Lam\'e coeficient  $\lambda\rho^s_0\to\lambda$, $\mu\rho^s_0\to\mu$ which define $\sigma^s$.
\end{remark}
\begin{proposition}\label{prop1}
Let $\gamma=\tr_{\vF\vF^T}$; then
\[
\gamma=\tr_{\vF\vF^T}=(2-2\n\cdot\vd + |\n\vd|^2)J^{2},~\tilde\gamma =\gamma J^{-2}
\]
and the following holds
\eq{\label{ab}&&
\sigma^s 
=\rho^s  \left(a \vI+2 b  (\D\vd-\n\vd\n^T\vd)\right),  ~~ with 
\cr&&
a=\lambda^s(\frac12\gamma-1)(\tilde\gamma -1)+ \mu^s(\gamma-J^2-1)\tilde\gamma,
\cr&&
b = \frac12(\frac{\lambda^s}2+\mu^s)(\gamma-1)-\frac{\lambda^s}4
}
\end{proposition}
{\it Proof}

First note that if $\vB=\vF\vF^T$ then
\eq{&&
\sigma^s = \rho^s\Big[[\lambda^s(\frac12\gamma-1)-\mu^s]\vB + \mu^s\vB^2\Big]
}
Now by the Cayley- Hamilton theorem in 2 dimensions, $\vB^2-\gamma\vB+J^2\vI=0$. As $\vB^{-1} = I -\D\vd +\n\vd\n^T\vd$  let $\vC=\vI-\vB^{-1}=\D\vd-\n\vd\n^T\vd$. Then
\eq{
\vB=\gamma\vI-J^2\vB^{-1}=(\gamma-J^2)\vI +J^2 \vC,
~
\vB^2=(\gamma^2-(1+\gamma)J^2)\vI+\gamma J^2\vC.~~
}
Therefore
\eq{&
\sigma^s&=\rho^s\Big[[\lambda^s(\frac12\gamma-1)-\mu^s][(\gamma-J^2)\vI +J^2 \vC]+\mu^s[(\gamma^2-(1+\gamma)J^2)\vI+\gamma J^2\vC]\Big]
\cr&&
=\rho^s\Big[[(\lambda^s(\frac12\gamma-1))(\gamma-J^2) 
\cr&&~~~~
+ \mu^s\gamma(\gamma-1-J^2)]\vI + [\lambda^s(\frac12\gamma-1)+\mu^s(\gamma-1) ]J^2\vC\Big]
}
\hfill{$\diamond$}

\subsection{Variational Monolithic Eulerian Formulation}
\emph{From now on we limit our analysis to the case $\rho_0^s$,$\rho_0^f$ constant.}

One must find $(\vu,p)$ with $\vu_{|\Gamma}=0$, $\vd$  and $\Omega_t^s$, $\Omega^f$, solution  for all $(\hat\vu,\hat p)$ with   $\hat\vu_{|\Gamma}=0$ of
 \eq{\label{monolc}\ds 
\left\{
\begin{array}{l}\ds
\int_{\Omega^f_t} \Big[\rho^f \Dt\vu\cdot\hat\vu -p\n\cdot\hat\vu -  \hat p\n\cdot\vu+
\frac{\mu^f}2\D\vu:\D\hat\vu \Big]
\cr\ds
+
\int_{\Omega^s_t} \rho^s \left[\Dt\vu\cdot\hat\vu +b(\D\vd-\n\vd\n^T\vd):\D\hat\vu
+a \n\cdot\hat\vu\right]
= \int_{\Omega_t} f\cdot\hat\vu
\cr
\Dt\vd=\vu,~~J^{-1}=\det_{\vI-\n\vd},~~\rho^r=J^{-1}\rho^r_0,
\cr
  \{\dot x(t)=\vu(x(t),t), ~x(0)=x_0\in\Omega_0^r~\Rightarrow~x(t)\in\Omega_t^r\}, ~r=s,f.
\end{array}\right.
}
For an existence result, up to time $T^*$, see \cite{boulakia} \cite{coutand} (see also \cite{vanni}),  provided a regularization term is added to the formulation to insure that $\p_t\vd$ has $H^1$-regularity; $T^*$ is such that the solid does not touch the boundary and $\Sigma_t$ does not buckle. 

\section{Numerical Scheme}
For the stability of the numerical scheme, the problem is that even for small displacements the Lam\'e terms $\mu^s\n\vu:\n\hat\vu$+$\lambda^s\n\cdot\vu\n\cdot\hat\vu$ are hidden in $b\D\vd:\D\hat\vu$ and $a\n\cdot\hat\vu$ in the above variational formulation (\ref{monolc}). 

But notice that
\eq{&&
J^2 =1+2\n\cdot\vd -2\det_{\n\vd} + 3(\n\cdot\vd)^2 + o(|\n\vd|^2)
\cr&&
\gamma=2(1+\n\cdot\vd + (\n\cdot\vd)^2 + \frac12|\n\vd|^2 - 2\det\n\vd) + o(|\n\vd|^2)
\cr&&
(\frac\gamma 2 - 1)(\tilde\gamma-1) =\n\cdot\vd - (\n\cdot\vd)^2 - \frac12|\n\vd|^2 - 2\det\n\vd + o(|\n\vd|^2)
}
So it makes sense to define 
\eq{\label{c}
c=a-\lambda^s\n\cdot\vd
}
To prepare the time discretisation of  (\ref{monolc}) with a given time step $\delta t$, let
\eq{\label{aaux}
\bar\vd=\vd -\delta t\vu
}
Then (\ref{monolc}) becomes
\[ 
 \left\{
\begin{array}{l}\ds 
\int_{\Omega^f_t} \Big[\rho \Dt\vu\cdot\hat\vu -p\n\cdot\hat\vu -  \hat p\n\cdot\vu+
\frac{\mu^f}2\D\vu:\D\hat\vu \Big]
\cr\ds
+
\int_{\Omega^s_t} \rho\delta t\Big[ b(\D\vu-\n\bar\vd\n^T\vu-\n\vu\n^T\bar\vd +\delta t\n\vu\n^T\vu):\D\hat\vu
+\lambda^s \n\cdot\vu~\n\cdot\hat\vu\Big]
\cr \ds
+\int_{\Omega^s_t} \rho\Big[ \Dt\vu\cdot\hat\vu +b(\D\bar\vd-\n\bar\vd\n^T\bar\vd):\D\hat\vu
+(c + \lambda^s\n\cdot\bar\vd)\n\cdot\hat\vu\Big]
= \int_{\Omega_t} f\cdot\hat\vu
\cr
\Dt\vd=\vu,~~\rho=\rho_0\det_{\vI-\n\vd}.
\end{array}\right.
\]
Here linear elasticity is visible because the zero order term of $b$ is $\frac{\mu^f}2$.  From now on we do not use $\bar\vd$ because the Characteristics-Galerkin discretisation of $\Dt\vd=\vu$ will give an analogue of (\ref{aaux}). 
\subsection{Discretisation of Total Derivatives}
Let $\Omega\subset\R^d$, $\vu\in\vH^1_0(\Omega)=(H^1_0(\Omega))^d$, ($d=2$ here), $t\in(0,T)$ and $x\in\Omega$. Then let $\chi^t_{\vu,x}(\tau)$ be the solution at time $\tau$ of
\[
\dot\chi(\tau)= \vu(\chi(\tau),\tau)\hbox{ with }\chi(t)=x.
\]
If $\vu$ is Lipschitz in space and continuous in time the solution exists.  The Characteristics-Galerkin method relies on the concept of total derivative:
\[
\Dt\vv(x,t):=\frac{d}{d\tau} \vv(\chi(\tau),\tau)|_{\tau=t}=\p_t\vv+\vu\cdot\n\vv.
\]
Given a time step $\delta t$, let us approximate 
\[
\chi^{(n+1)\delta t}_{\vu^{n+1},x}(n\delta t)\approx \Y^{n+1}(x):= x - \vu^{n+1}(x)\delta t
\]
\begin{remark}
Note also that, as $J\rho$ is convected by $\vu$, that is $J\rho|_{\chi^t_{\vu,x}(\tau),\tau}=J\rho|_{x,t}$, so
a consistent approximation is
\[
(J^n\rho_n)\circ\Y^{n+1}(x)=J^{n+1}(x)\rho_{n+1}(x),~x\in\Omega_{n+1}.
\]
Thus discretizing the total derivative of $\vu$ or the one of $\rho_0\vu$ will give the same scheme.
\eq{&&\ds
\rho_0(x)\frac{\vu^{n+1}(x)-\vu^n(\Y^{n+1}(x))}{\delta t}=J^{n+1}\rho_{n+1}\frac{\vu^{n+1}-\vu^n\circ\Y^{n+1}}{\delta t}
\cr&&\ds
=\frac{J^{n+1}\rho_{n+1}\vu^{n+1}-(J^n\rho_n\vu^n)\circ\Y^{n+1}}{\delta t}
= \frac{\rho_{0}\vu^{n+1}-(\rho_0\vu^n)\circ\Y^{n+1}}{\delta t}
}
\end{remark}
\subsection{Updating the fluid and solid domain}
From the definition of $\Y$, notice that the only way to be consistent is to define $\Omega_{n+1}$ using $\vu^{n+1}$, i.e. implicitly, since the later is defined also on $\Omega_{n+1}$:
\[
\Omega_{n+1}=(\Y^{n+1})^{-1}(\Omega_n)=\{x:~\Y^{n+1}(x):=x-\vu^{n+1}(x)\delta t\in\Omega_n\}
\]
\subsection{The Time Discretized Scheme}
 Let
\eq{\label{dscheme}
\tilde\vd^n:=\vd^n\circ\Y^{n+1}
,~\vd^{n+1}=\tilde\vd^n+\delta t\vu^{n+1},~\rho_{n+1}=\rho_0 \det_{\vI-\n\vd^{n+1}}.
}
Let $\tilde b_n, \tilde c_n$ be given by (\ref{ab},\ref{c}) computed with $\tilde d^n$.
The following defines $\vu^{n+1},p^{n+1}$ with $\vu^{n+1}|_{\Gamma}=0$: $\forall \hat\vu,\hat p$, with $\hat\vu|_{\Gamma}=0$,
 \eq{\label{monolfulldt}
\left\{ \begin{array}{l}\ds
\int_{\Omega_{n+1}} \rho_{n+1}\frac{\vu^{n+1}-\vu^n\circ\Y^{n+1}}{\delta t}\cdot\hat\vu 
\cr\ds
~~~+ \int_{\Omega^f_{n+1}}\Big[ -p^{n+1}\n\cdot\hat\vu -  \hat p\n\cdot\vu^{n+1}+
\frac{\mu^f}2\D\vu^{n+1}:\D\hat\vu \Big]
\cr\ds
+
\int_{\Omega^s_{n+1}} \rho_{n+1}\delta t\Big[ \tilde b_{n}(\D\vu^{n+1}-\n\tilde\vd^n\n^T\vu^{n+1}-\n\vu^{n+1}\n^T\tilde\vd^n ):\D\hat\vu
\cr\ds
\hskip 2cm+\lambda^s \n\cdot\vu^{n+1}\n\cdot\hat\vu
\Big]
\cr \ds
+\int_{\Omega^s_{n+1}} \Big[ \tilde b_{n}(\D\tilde\vd^n-\n\tilde\vd^n\n^T\tilde\vd^n):\D\hat\vu
+(\tilde c_{n} + \lambda^s\n\cdot\tilde\vd^n)\n\cdot\hat\vu\Big]
\cr\ds
= \int_{\Omega_{n+1}} f\cdot\hat\vu.
\end{array}\right.
}
\subsection{Iterative Solution by Fixed Point}
The most natural method to solve the above is to freeze some coefficients so as to obtain a well posed linear problem and iterate:
\begin{enumerate}
\item Start with $\vu=\vu^n$,  $\Y(x)=x-\vu\delta t$, $\Omega^r=\Y^{-1}(\Omega^r_n)$, $r=s,f$.
\item Set $\tilde\vd^n=\vd^n\circ\Y$, $\tilde\rho_n=\rho_0\det_{\vI-\n\tilde\vd}$; compute $\tilde b_n,\tilde c_n$.
\item Find $\vu^{n+1},p^{n+1}$ by solving
 \eq{\label{iter}
 \left\{\begin{array}{l}
 \ds
\int_{\Omega} \tilde\rho_n\frac{\vu^{n+1}-\vu^n\circ\Y}{\delta t}\cdot\hat\vu 
\cr\ds
~~~+ \int_{\Omega^f}\Big[ -p^{n+1}\n\cdot\hat\vu -  \hat p\n\cdot\vu^{n+1}+
\frac{\mu^f}2\D\vu^{n+1}:\D\hat\vu \Big]
\cr\ds
+
\int_{\Omega^s}\tilde \rho_n\delta t\Big[ \tilde b_n(\D\vu^{n+1}-\n\tilde\vd^n\n^T\vu^{n+1}-\n\vu^{n+1}\n^T\tilde\vd^n ):\D\hat\vu
\cr\ds
\hskip 2cm+\lambda^s \n\cdot\vu^{n+1}\n\cdot\hat\vu 
\Big]
\cr\ds
+\int_{\Omega^s} \Big[ \tilde b_n(\D\tilde\vd^n-\n\tilde\vd^n\n^T\tilde\vd^n):\D\hat\vu
+(\tilde c_n + \lambda^s\n\cdot\tilde\vd^n)\n\cdot\hat\vu\Big]
\cr\ds
= \int_{\Omega} f\cdot\hat\vu
\end{array}\right.
}
\item Set $\vu=\vu^{n+1}, \Y(x)=x-\vu\delta t,\Omega^r=\Y^{-1}(\Omega^r_n),~r=s,f$.
\item If not converged return to Step 2 else set $\vd^{n+1}=\vd^n\circ\Y+\delta t\vu^{n+1}$.
\end{enumerate}
Notice that (\ref{iter}) is a well posed linear problem whenever 
\[
A(\vu,\hat\vu)=\int_{\Omega^s}\left[\frac{\rho}{\delta t}\vu\cdot\hat\vu+\tilde b(\D\vu-\n\tilde\vd^n\n^T\vu-\n\vu\n^T\tilde\vd^n ):\D\hat\vu
+\lambda^s \n\cdot\vu\n\cdot\hat\vu\right]
\]
is coercive. Then (\ref{iter})  gives a solution bounded in $\vH^1(\Omega)$ and converging subsequences can be extracted from 
$\rho_{n+1},\vu^{n+1},\Omega^r_{n+1}$ when $\bar\Omega=\overline{\Omega^f_n\cup\Omega^s_n}$ is fixed. Then convergence would occur if we could prove that $\Omega^r_{n+1}$ converges.
\subsection{Spatial Discretisation with Finite Elements}
Let $\T_h^0$ be a triangulation of the initial domain. Spatial discretisation can be done with the most popular finite element for fluids: the Lagrangian triangular elements of degree 2 for the space $V_h$ of velocities and displacements and Lagrangian triangular elements of degree 1 for the pressure space $Q_h$; later we will also discuss the stabilised $P^1-P^1$ element;  provision must be made for two pressure variables, one  in the structure and one in the fluid because the pressure is discontinuous at the interface $\Sigma$; therefore $Q_h$ is the space of piecewise linear functions on the triangulation continuous in $\Omega^r_{n+1},~r=s,f$. A small penalization with parameter $\epsilon$  must be added to impose uniqueness of the pressure. 

This leads us to find $\vu^{n+1}_h\in V_{h0_\Gamma}$, $p^{n+1}_h\in Q_h$, $\Omega_{n+1}$
such that for all $\hat\vu_h,\hat p_h\in V_{h0_\Gamma}\times Q_h$ with
\[
 \tilde\vd_h^n:=\vd_h^n\circ\Y^{n+1},~\hbox{ where }
\Y^{n+1}(x)=x-\vu^{n+1}_h(x)\delta t,
\]
the following holds:
 \eq{\label{monolfulldt0}
 \left\{
 \begin{array}{l}\ds 
{\bf a}(\tilde\rho_n,\tilde b_n,\tilde c_n;\vu^{n+1},\hat\vu):=\int_{\Omega_{n+1}} \tilde\rho_{n}\frac{\vu_h^{n+1}-\vu_h^n\circ\Y^{n+1}}{\delta t}\cdot\hat\vu_h 
\cr\ds
~~~+ \int_{\Omega^f_{n+1}}\Big[ -p^{n+1}\n\cdot\hat\vu_h -  \hat p\n\cdot\vu_h^{n+1}+
\frac{\mu^f}2\D\vu_h^{n+1}:\D\hat\vu_h \Big]
\cr\ds
+
\int_{\Omega^s_{n+1}} \tilde\rho^s_{n}\delta t\Big[ \tilde b_{n}(\D\vu_h^{n+1}-\n\tilde\vd_h^n\n^T\vu_h^{n+1}-\n\vu_h^{n+1}\n^T\tilde\vd_h^n ):\D\hat\vu_h
\cr\ds
\hskip 2cm+\lambda^s \n\cdot\vu_h^{n+1}\n\cdot\hat\vu_h 
\Big]
\cr\ds 
+\int_{\Omega^s_{n+1}} \Big[ \tilde b_{n}(\D\tilde\vd_h^n-\n\tilde\vd_h^n\n^T\tilde\vd_h^n):\D\hat\vu_h
+(\tilde c_{n} + \lambda^s\n\cdot\tilde\vd_h^n)\n\cdot\hat\vu_h\Big]
\cr\ds
= \int_{\Omega_{n+1}} \vf\cdot\hat\vu_h,~~~~~~ 
\Omega_{n+1}=(\Y^{n+1})^{-1}(\Omega_n)=\{x:~\Y^{n+1}(x)\in\Omega_n\}.
\end{array}\right.}
Then $\vd$ 
\[
\vd_h^{n+1}=\tilde\vd_h^n+\delta t\vu_h^{n+1},
\]
\subsection{Implementation}
The various tests we made lead us to recommend the following:
\begin{itemize}
\item Move the vertices of the mesh in the structure with its own velocity:
\eq{\label{move}
q^{n+1}_i=q^n_i+\vu_h^{n+1}(q^{n+1}_i)\delta t
} 
which, as explained above has to be implemented through an iterative process.
\item Remesh the fluid part at each iteration with a Delaunay-Voronoi mesh generator from the boundary vertices ($\Sigma_{n+1}$ included).

This required the development of a specific module to identify computationally the vertices of the fluid-structure interface $\Sigma$, which are then input to the fluid mesh generator.

\item In doing so, the discrete topological properties of the structural part are preserved and we have the  important property that the value $\vd[i]$ of $\vd$ at  vertex $q_i$ in the computer implementation of $\vd$ by an array of values at the nodes, satisfies
\[
\vd^{n+1}[i]=\vd^n[i] + \delta t\vu^{n+1}[i],~~\forall i.
\]
In other words $\vd^n\circ\Y^{n+1}$ is $\vd^n[i]$ after moving the vertices by (\ref{move}).
\end{itemize}

\section{Energy Estimate}

\subsection{Stability of the Scheme Discretized in Time}
To conserve energy we need to change the scheme (\ref{monolfulldt0}) slightly,  from
\eq{\label{monolfulldt2}&&
{\bf a}(\tilde\rho_n,\tilde b_n,\tilde c_n;\vu^{n+1},\hat\vu) = \int_{\Omega_{n+1}} \vf\cdot\hat\vu_h~\hbox{ to }
\cr&&
{\bf a}( \rho_{n+1},b_{n+1}, c_{n+1};\vu^{n+1},\hat\vu) 
+ \delta t^2 \int_{\Omega^s_{n+1}} \rho^s_{n+1}b_{n+1}\n\vu_h^{n+1}\n^T\vu_h^{n+1}:\D\hat\vu_h~~
\cr&&
= \int_{\Omega_{n+1}} \vf\cdot\hat\vu_h
}
\begin{lemma}\label{prop2}
The mapping $\vX^n:\Omega_0\mapsto\Omega_n$ is also $\vX^{n+1}=(\Y^{n+1})^{-1}\circ\vX^n$, $n\geq 1$ and 
the jacobian of the transformation is $\vF^n:=\n^T_{x_0}\vX^n=(\vI-\n\vd^n)^{-T}$.
\end{lemma}
\emph{Proof}

Notice that 
$
\Y^1(\Y^{2}(..\Y^{n-1}(\Y^n(\Omega_n))..))=\Omega_0
$
Hence 
\[
\vX^{n+1}=[\Y^1(\Y^{2}(..\Y^{n}(\Y^{n+1})))]^{-1}=(\Y^{n+1})^{-1}\circ\vX^n .
\]
By definition of $\vd^{n+1}$ in (\ref{dscheme})
\eq{\label{eq3}&
\vd^{n+1}(\vX^{n+1}(x_0))&=\vd^n(\Y^{n+1}(\vX^{n+1}(x_0)))+\vu^{n+1}(\vX^{n+1}(x_0))\delta t
\cr&&
=\vd^n(\vX^{n}(x_0))+\vu^{n+1}(\vX^{n+1}(x_0))\delta t,
}
and since 
$
\vX^{n+1}(x_0)=\vd^{n+1}(\vX^{n+1}(x_0))+x_0$ we have that
\eq{&
{\vF^{n+1}}&={\n^t_{x_0}}(\vd^{n+1}((\vX^{n+1}(x_0)))+x_0),
\cr&&
={\n\vd^{n+1}}^T\vF^{n+1}+\vI~\Rightarrow~\vF^{n+1}=(\vI-\n\vd^{n+1})^{-T}
}
\hfill{$\diamond$}

Note that (\ref{eq3}) shows also that
\eq{\label{eq5}
\vF^{n+1} =\vF^n + \delta t\n_{x_0}^T\vu^{n+1}
}
\begin{lemma}
With $\Psi$ defined by (\ref{defpsi}),
\eq{& \ds
\int_{\Omega^s_{n+1}} \rho^s_{n+1} &\left[b^{n+1}(\D\vd^{n+1}-\n\vd^{n+1}\n^T\vd^{n+1}):\D\hat\vu
+a^{n+1} \n\cdot\hat\vu\right]
\cr&&
=\int_{\Omega^s_0}\p_{\vF}\Psi^{n+1}:\n_{x_0}\hat\vu
}
\end{lemma}
\emph{Proof}
By Proposition  \ref{prop1} and Lemma \ref{prop2}:
\eq{&& \ds
\int_{\Omega^s_{n+1}}\rho^s_{n+1} \left(a^{n+1} \vI+2 b^{n+1}  (\D\vd^{n+1}-\n\vd^{n+1}\n^T\vd^{n+1})\right):\n\hat\vu 
\cr&&
=\int_{\Omega^s_{n+1}}\sigma^s_{n+1}:\n\hat\vu 
=\int_{\Omega^s_{n+1}}\left[\rho^s_{n+1}\p_{\vF}\Psi\vF^T\right]|_{n+1}:\n\hat\vu 
\cr&&
=\int_{\Omega^s_{n+1}}\left[J_{n+1}^{-1}\rho^s_0\p_{\vF}\Psi\vF^T\right]|_{n+1}:\n\hat\vu 
=\int_{\Omega^s_0}\rho^s_0\p_{\vF}\Psi^{n+1}:\n\hat\vu 
}
\hfill{$\diamond$}
\begin{theorem}
When $f=0$ and $\rho_0$ is constant in each domain $\Omega^r_0,~r=s,f$, the numerical scheme (\ref{monolfulldt2}) has the following property:
\eq{\label{monolfulldt3}
\int_{\Omega_{n}}\frac{\rho_{n}}2|\vu^{n}|^2
 + \delta t\sum_{k=1}^n\int_{\Omega_k^f}\frac{\mu^f}{2}|\D\vu^{k}|^2 
+ \int_{\Omega^s_0}\Psi^{n}
 \leq \int_{\Omega_{0}}\frac{\rho_{0}}2|\vu^{0}|^2+\int_{\Omega^s_0}\Psi^{0}
}
\end{theorem}
\emph{Proof}
Let $r=s$ or $f$.
Let us choose $\hat\vu=\vu^{n+1}$ in (\ref{monolfulldt2}).  By Schwartz inequality
\begin{eqnarray*}&&
\int_{\Omega^r_{n+1}}\rho_{n+1}(\vu^{n}\circ\Y^{n+1})\cdot\vu^{n+1}
=
\rho_0\int_{\Omega^r_{n+1}}(J^{n+1})^{-1}(\vu^{n})\circ\Y^{n+1})\cdot\vu^{n+1}
\cr&&
\leq 
\rho_0\left(\int_{\Omega^r_{n+1}}(J^{n+1})^{-1}(\vu^{n}\circ\Y^{n+1})^2\right)^{\frac12}
\left(\int_{\Omega^r_{n+1}}(J^{n+1})^{-1}(\vu^{n+1})^2\right)^{\frac12}
\cr&&
=\left[\int_{\Omega^r_{n}}\rho_n(\vu^{n})^2
\int_{\Omega^r_{n+1}}\rho_{n+1}(\vu^{n+1})^2\right]^{\frac12}
\leq\frac12 \int_{\Omega^r_{n}}\rho^r_n{\vu^{n}}^2+ \frac12\int_{\Omega^r_{n+1}}\rho^r_{n+1}{\vu^{n+1}}^2,
\end{eqnarray*}
Plugging this estimate in (\ref{monolfulldt2}) with $\hat\vu=\vu^{n+1}$ leads to
{\small\[\ds
\int_{\Omega_{n+1}}
\frac{\rho_{n+1}}2|\vu^{n+1}|^2
+\delta t\int_{\Omega^f_{n+1}}\frac{\mu^f}{2}|\D\vu^{n+1}|^2
+\int_{\Omega_0}\Psi^{n+1}
\leq 
\int_{\Omega_{n}}\frac{\rho_{n}}{2}|\vu^{n}|^2
+\int_{\Omega_0}\Psi^n
\]}
\hfill{$\diamond$}

\subsection{Energy Estimate for the Fully Discrete Scheme}
The proof for the spatially continuous case will work for the discrete case if 
\eq{\label{compose}
\vX^{n}=\vX^{n+1}\circ\Y^{n+1}.
}
As discussed in \cite{fhopfsi} it may be possible to program an isoparametric  $P^2-P^1$ element for which (\ref{compose})  but it is certainly far from easy.
On the other hand, consider the stabilised $P^1-P^1$ element: the fluid pressure and the solid pressure are continuous and piecewise linear on the triangulation. The inf-sup condition for stability does not hold unless the incompressibility  
condition in the fluid, $\n\cdot\vu=0$, is changed to $-\epsilon\Delta p+\n\cdot\vu=0$, (see \cite{boffibrezzifortin} for details). It amounts to adding $\epsilon\n p^{n+1}\n\hat p$ next to the term with $\mu^f$ in the variational formulations.
\begin{figure}[ht]
	\begin{minipage}[b]{0.45\linewidth}
	\centering
\begin{tikzpicture}[xscale=0.8,yscale=0.8]
\draw [black] (8,-2) to (10,-2) to (10,0) to (8,-2) ;
\draw [fill] (8,-2) circle [radius=0.05];
\draw [fill] (10,-2) circle [radius=0.05];
\draw [fill] (10,0) circle [radius=0.05];
\node at (9.6,-1.05) {$T^k_0$};
\draw [black] (8,1) to (10,0.5) to (10,2) to (8,1) ;
\draw [fill] (8,1) circle [radius=0.05];
\draw [fill] (10,0.5) circle [radius=0.05];
\draw [fill] (10,2) circle [radius=0.05];
\node at (9.5,1) {$T^k_n$};
\draw [black] (8+3,0) to (10+3,0.5) to (10+1,2) to (8+3,0) ;
\draw [fill] (8+3,0) circle [radius=0.05];
\draw [fill] (10+3,0.5) circle [radius=0.05];
\draw [fill] (10+1,2) circle [radius=0.05];
\node at (9+2.5,1) {$T^k_{n+1}$};
\draw[thick,->](8.75,-1)to(8.75,0.9);
\node at (8.2,0) {$\vX^{n}$};
\draw [thick,<-] (9.9,1.2) to (10.9,1.2);
\node at (10.55,1.5) {$\Y^{n+1}$};
\draw[thick,->](10.1,-1)to(11.1,0.8);
\node at (11.0,-0.5) {$\vX^{n+1}$};
\node at (9.8,2.3) {$q^j_{n}$};
\node at (11.2,2.3) {$q^j_{n+1}$};
\end{tikzpicture}
	\end{minipage}
	\begin{minipage}[b]{0.45\linewidth}
	\centering
\caption{\label{triangles}{\it { Sketch to understand if $\vX^n=\Y^{n+1}\circ\vX^{n+1}$ holds with the $P^1-P1$ stabilised element. A  triangle $T^k_0$ in the reference domain (chosen here to be its initial position at time zero)  becomes   triangle $T^k_n$ at $t_n$ and $T^k_{n+1}$ at time $t_{n+1}$:  $T^k_n=\vX^n(T^k_0)$ and $T^k_{n+1}=\vX^{n+1}(T^k_0)$ Vertices are preserved by these transformations.}}}
\end{minipage}
\end{figure}
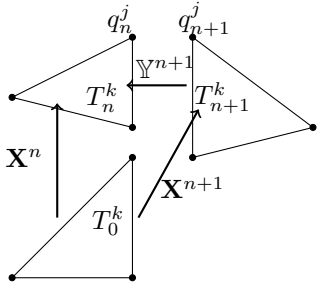
Then (\ref{compose}) holds (see Figue \ref{triangles}) and the proof of the spatially continuous case can be adapted leading to (\ref{monolfulldt3}) with an additional viscous term $\epsilon|\n p^{n+1}|^2$ next to the term with $\mu^f$.

\begin{remark} 
Because of energy preservation scheme (\ref{monolfulldt2}), implemented via a fixed point algorithm as in (\ref{iter}), generates bounded sequences $\rho,\vu,q^i$; it seems safe to assess that out of these bounded subsequences will converge to a solution of the problem discretized in space but continuous in time when $\delta t\to 0$.
\end{remark}

\section{Numerical Tests}
In our tests we have used the $P^2-P^1$ element, confident that it will behave as well as the stabilised $P^1-P1$ element as indicated in \cite{fhopfsi}.

\subsection{The Cylinder-Flag Test}
A compressible hyperelastic Mooney-Rivlin material, shaped as a rectangle of size $[0,l]\times[0,h]$, is attached behind a cylinder of radius $r$ and beats in tune with the Karman vortices of the wake behind the cylinder; the fluid in the computational rectangular domain $[0,L]\times[0,H]$ enters from the left and is free to leave on the right.  The center of the cylinder is at $(c,c)$ (see figure \ref{sketch}). In \cite{dunne} the following numerical values are suggested:
\begin{description}
\item[Geometry] $l=0.35$, $h=0.02$, $L=2.5$, $H=0.41$, $c=0.2$ which puts the cylinder slightly below the symmetry line. 
\item[Fluid]  density $\rho^f=10^3 kg/m^{3}$ and a reduced viscosity $\ds\nu^f=\frac{\mu^f}{\rho^f}=10^{-3}m^2/s$; inflow horizontal velocity $\ds\vu(0,y)= \bar U\left(\frac{6}{H^2}y(H-y),0\right)^T$ is a parabolic profile with flux $\bar U H$. Top and bottom boundaries are walls with no-slip conditions.
\item[Solid] $\ds E = 2\mu(1+\sigma)/\rho^s $, $\sigma = 0.4$, $\ds\lambda = \frac{E\sigma}{(1+\sigma)(1-2\sigma)}$.
\end{description}
Initial velocities and displacements are zero.  In all cases the same mesh is used initially with 2500 vertices. The time step is 0.005.

\subsubsection{Free Fall of a Thick Flag}
The gravity is $g=9.81$ in $\Omega_t$. When $\bar U=0$, $\mu=0.135 10^{6}$ and $\rho^s=20\rho^f$, the flag falls under its own weight; it comes to touch the lower boundary with zero velocity at time 0.49 and then moves up under its spring effect. This test is named FLUSTRUK-FSI-2$^*$ in \cite{dunne} but we have used a different value for $\mu$ because the one reported in \cite{dunne} seems unlikely. 

\begin{figure}[htbp]
\begin{center}
\includegraphics[width=9cm]{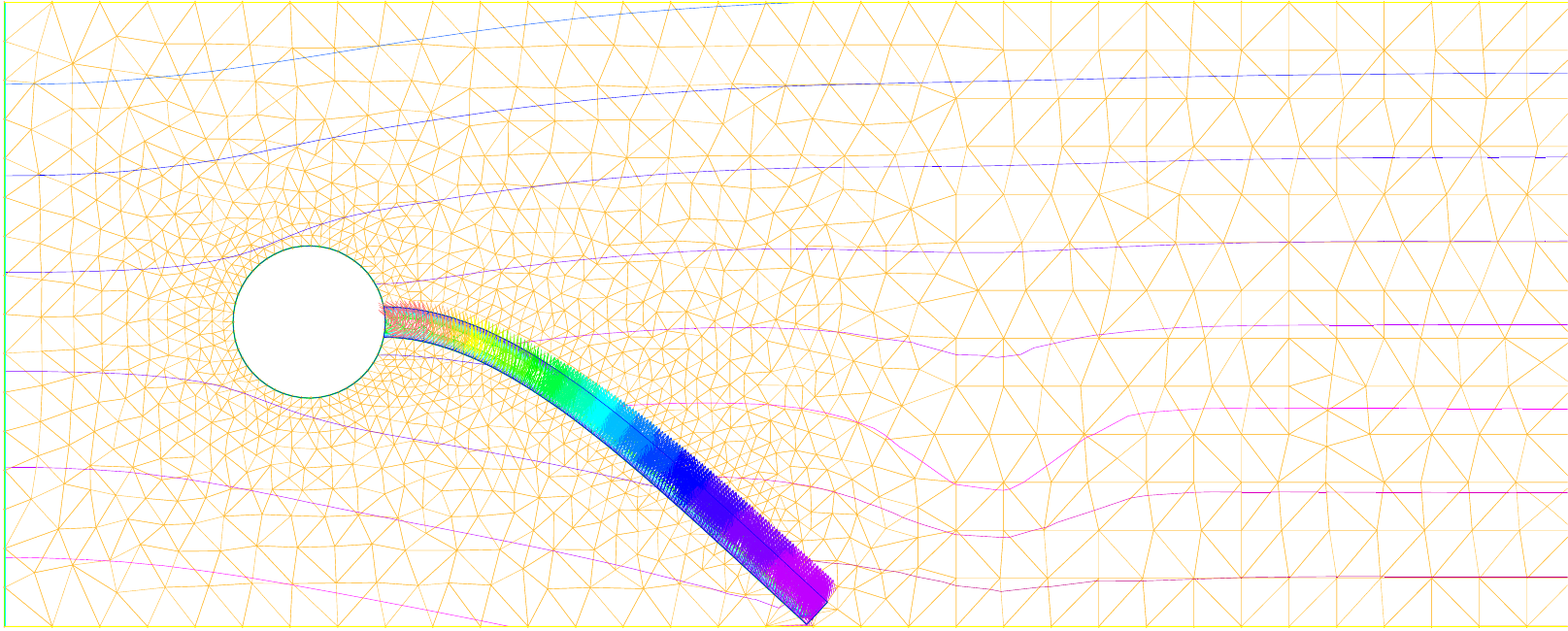}
\caption{\label{turek1a}{\it FLUSTRUK-FSI-2$^*$\cite{dunne}.  Zoom near the flag at t=0.495 just as it begins to move up after the fall under its own weight in a flow initially at rest.Mesh and Pressure lines are shown in the fluid and velocity vectors in the solid.}}
\includegraphics[width=5cm]{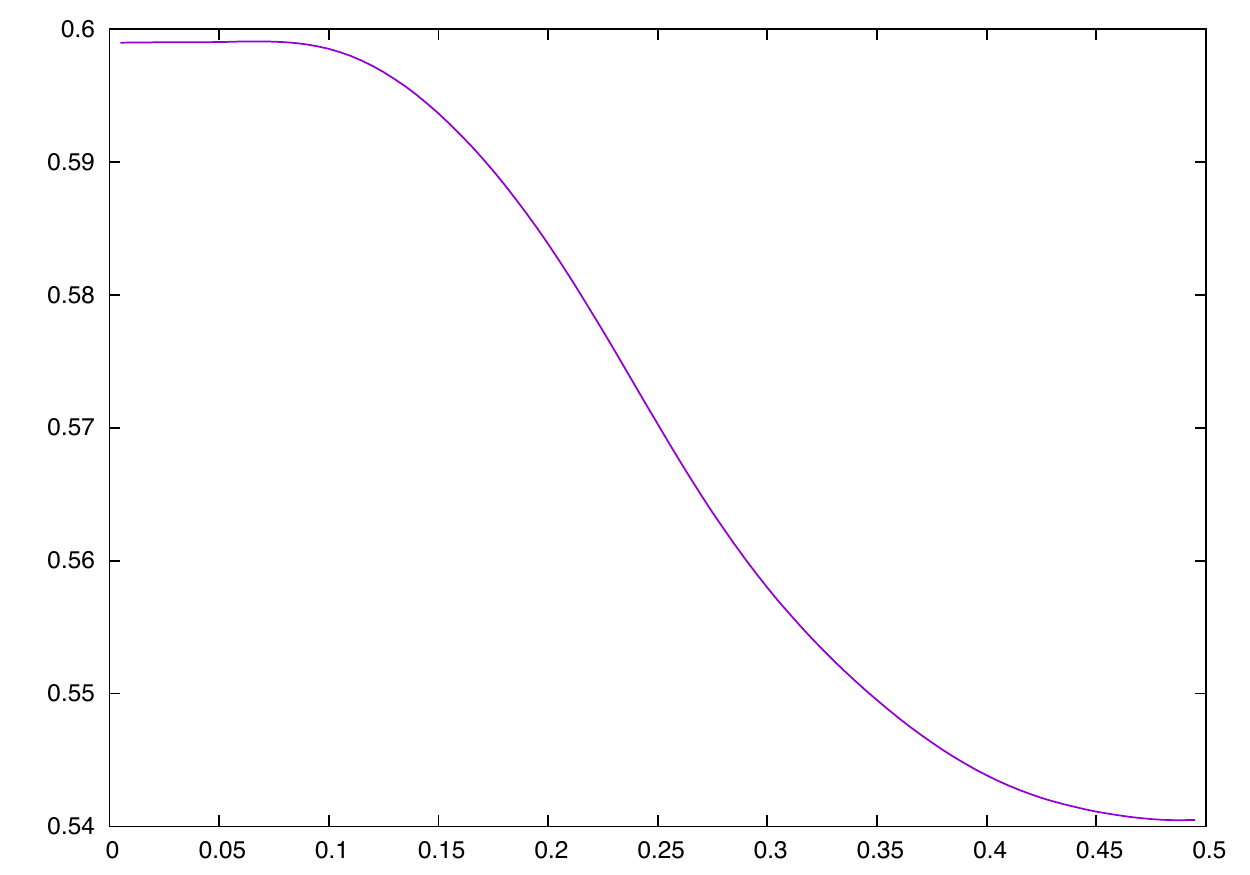}
\includegraphics[width=5cm]{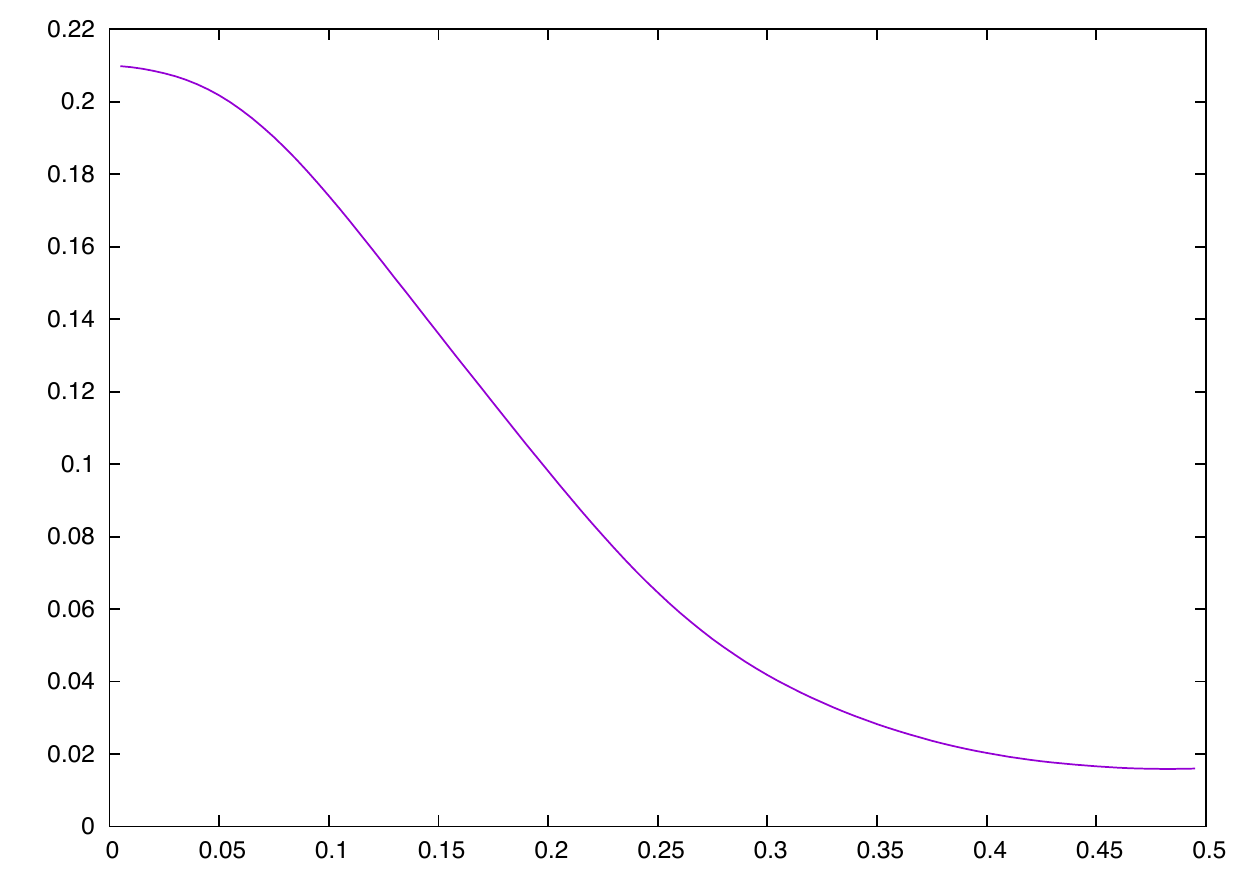}
\caption{\label{turek1b}{\it FLUSTRUK-FSI-2$^*$ test of \cite{dunne}. Position of the upper right corner of the flag versus time: x vs t on the left and y vs t on the right. }}
\end{center}
\end{figure}

Figure \ref{turek1a} shows a zoom around the flag at the time when it has stopped to descend and  started to move upward. 
Pressure lines are drawn in the flow region together with the mesh and the velocity vectors in the flag and drawn at each vertex. Figure \ref{turek1b}  shows the coordinates of the upper right tip of the flag versus time.

\subsubsection{Flow past a Cylinder with a Thick Flag Attached}
\begin{figure}[htbp]
\begin{center}
\includegraphics[width=12cm]{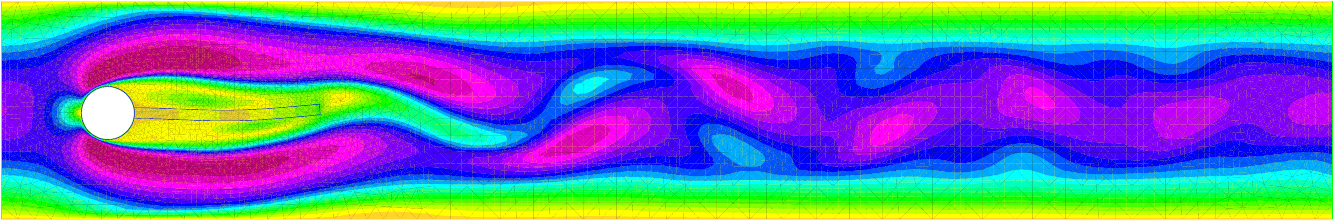}
\caption{\label{turek2b}{\it FLUSTRUK-FSI-3 Test. Color map based on the norm of the fluid and solid velocity vectors}}
\end{center}
\end{figure}

\begin{figure}[htbp]
\begin{center}
\includegraphics[height=5cm,width=12cm]{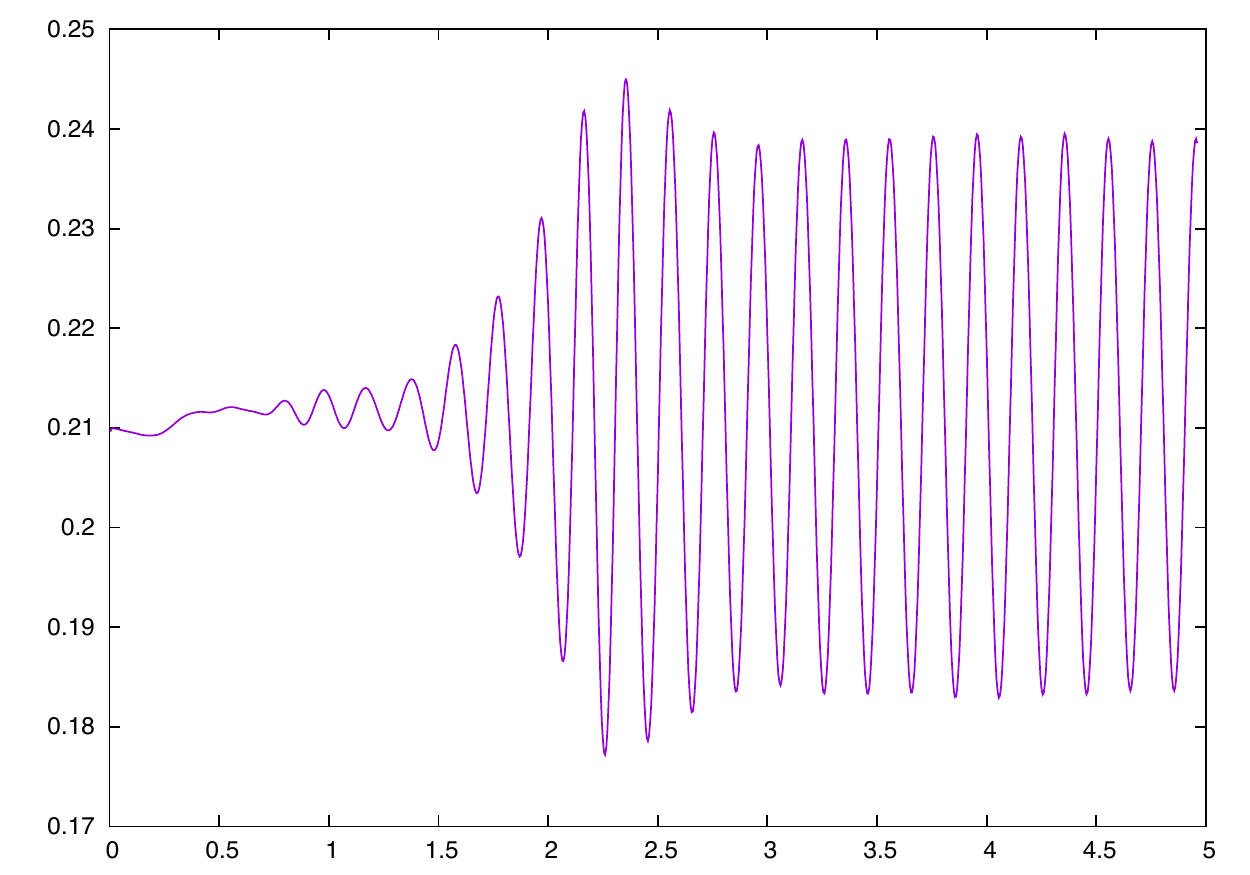}
\caption{\label{turek2c}{\it FLUSTRUK-FSI-3 Test.  Vertical position of the upper right tip of the flag versus time shown up to t=5}}
\end{center}
\end{figure}

This test is known as FLUSTRUK-FSI-3 in \cite{dunne}.  The geometry is the same as above but now $\bar U=2$, $\mu=2 10^6$ and $\rho^s=\rho^f$.  After some time a Karman-Vortex alley develops and the flag beats accordingly.  Results are shown on Figures \ref{turek2b} and \ref{turek2c}; the first one displays a snapshot of the velocity vector norms and the second the y-coordinate versus time of the top right corner of the flag.

These numerical results compare reasonably well with those of \cite{dunne}. The frequency is $5s^{-1}$ compared to $5.04$ and the maximum amplitude 0.031 compared to 0.032. However the results are sensitive to the time step.

\section*{Conclusion}
A fully Eulerian fluid-structure formulation has been presented for compressible materials with large displacements, discretized by an implicit first order Euler Scheme and the $P^2-P^1$ or stabilised $P^1-P^1$ elements. An energy estimate has been obtained which guarantees the stability of the scheme so long as the motion of the vertices does not flip-over a triangle. The method has been implemented with \texttt{FreeFem++}\cite{freefem}.  It is reasonably robust when the vertices in the structure are moved by their velocities and the fluid is remeshed with an automatic Delaunay mesh generator.  The method is first order in time and therefore somewhat too diffusive for delicate tests.  It is being extended to 3D and to second order in time discretisation.

\subsection*{Acknowledgement} The author thanks Fr\'ed\'eric Hecht for very valuable discussions and comments.

\end{document}